\documentclass{iopart}
\usepackage{graphicx}
\usepackage{latexsym}
\usepackage{amssymb}
\begin{document}

\title[Spin-polarized electron transport in (GaSb)$_2$M (M=Cr,Mn)]{Spin-polarized electron transport in the high-pressure ferromagnetic phases (GaSb)$_2$M (M=Cr,Mn)}
\author{A A Pronin$^2$, M V Kondrin$^1$, V R Gizatullin$^1$ , O A Sazanova$^1$,  A G Lyapin$^1$, S V Popova$^1$}
\address{$^1$ Institute for High Pressure Physics RAS, Troitsk, 142190 Russia}
\address{$^2$ General Physics Institute RAS, 117942 Moscow, Russia}

\begin{abstract}
For the first time magnetrotransport of the ferromagnetic high-pressure phases of (GaSb)$_2$M (M=Cr,Mn) was measured. It was found that the main component of magnetotransport in these phases is negative and its amplitude is increasing with temperature rising and approaching Curie temperature. Measuring of magnetoresistance in the cycling magnetic fields demonstrated that the Yosida component of magnetresistance is negligible and the main part of the magnetoresistance can be attributed to the spin-polarized electron transport.
\end{abstract}
\section{Introduction}

\begin{figure}
\includegraphics[width=0.7\textwidth]{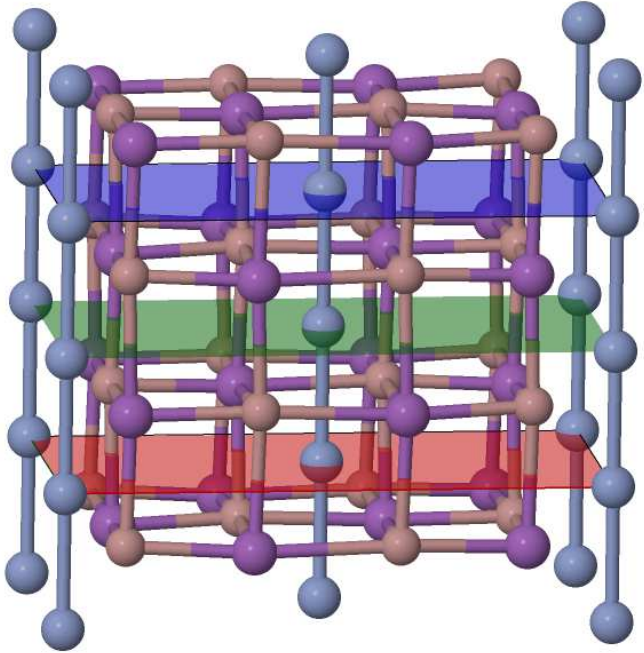}
\caption{Schematic stick-and-ball structure of orthorombic (GaSb)$_2$M (M=Cr,Mn) phases. Grey symbols -- magnetic ions; magenta and brown symbols -- distorted 6-coordinated GaSb lattice. The bottom and top translucent planes are the boundaries of the conventional unit cell.}
\label{fig:struct}
\end{figure}

There is a number of materials obtained by high-pressure treatment of semiconductors with sphalerite/diamond lattice and metals (Ge$_4$Mn \cite{takizawa:jssc90,sakakibara:jcsj09}, (GaSb)$_2$Mn\cite{en*popova:ftt06,en*kondrin:jetpl06,kondrin:jopcs08,sakakibara:jcsj09}, (GaSb)$_2$Cr\cite{sakakibara:jac10,kondrin:jopcm11}) which posses a set of properties (ferromagnetism and high almost semiconductor-like electrical resistivity) interesting from the point of view of spintronic applications\cite{dietl:08}. The high-pressure transitions leading to the synthesis of these materials can be described as transformation of  4-coordinated diamond-like lattice to 6-coordinated at higher pressures and certain ordering of metal atoms at intersticial cites accompanied with a slight distortion of the lattice which leads to the doubling of one lattice parameters (see Fig.~\ref{fig:struct}).

Transformation to the  6-coordinated crystal lattice  (which can be roughly approximated by $\beta$-Sn lattice, see e.g. \cite{mujica:rmp03,fedotov-barkalov:jopcm09}) at elevated pressures is commonly observed for substances with diamond-like structures, like pure GaSb. However, the transition leading to formation of (GaSb)$_2$M phases is quite different. The cell duplication in this case results in orthorhombic cell with one lattice parameter almost twice as large as two others. In formal way the structure of (GaSb)$_2$Cr (and similar to it structure of (GaSb)$_2$Mn) can be described in $Iba2$ space group with 4 structural units in conventional unit cell (Z=4) and lattice parameters  $a$=11.77, $b$=5.96, $c$=5.87\AA \cite{sakakibara:jac10,kondrin:jopcm11}. Interesting to note, that in the phase diagram of Cr-Si, for example, there is no thermodynamically stable phases with composition 1:4 \cite{goldschmidt:jlcm61}, similar to the (GaSb)$_2$M, so these compounds recovered from the high pressure are likely to be metastable at the ambient pressure.

It was previously shown that both these phases (GaSb)$_2$Mn and (GaSb)$_2$Cr are ferromagnetic with $T_C \approx$ 300 and 360 K respectively\cite{en*popova:ftt06,en*kondrin:jetpl06,kondrin:jopcs08,sakakibara:jcsj09,sakakibara:jac10,kondrin:jopcm11}. In the former case Curie temperature significantly depends on Mn content \cite{en*kondrin:jetpl06}, so small deviation from stochiometry can lead to significant lowering of the critical temperature, the effect which is not observed in (GaSb)$_2$Cr. Both materials demonstrates high level of resistivity ($\rho \approx$ 0.01 Ohm cm similar to highly doped semiconductors) with very weak dependence on temperature but with a quite pronounced peak at the vicinity of Curie temperature\cite{en*kondrin:jetpl06,kondrin:jopcm11}. This allows one to assume that there is certain correlation between zero-field resistivity and the magnetic properties caused by spin-polarization of electron transport. 

However, the first-principle calculation \cite{kulatov:ijqc13,kulatov:epl11,magnitskaya:pssc14} demonstrated that there is a small concentration of carriers at Fermi-level. In terms of minimal metallic conductivity this concentration is such as if every unit cell contributes about 0.1 $e$ to the conduction band. Computer simulations also predicts appearances of  two sharp peaks of electron density in the vicinity of Fermi-level,  so one may conclude that though these materials are metals but the additional density of states in the vicinity of Fermi level can lead to certain polarization of electron transport in them. Taking into account the ordering of metal ions along one direction (see Fig.~\ref{fig:struct}) one may suppose a significant anisotropy of electron transport properties but calculations disproves this: the ratio of conductivity along different crystallographic directions does not exceed 1.5\cite{kulatov:ijqc13,kulatov:epl11}. 

To clarify the point of electron transport polarization in this paper we present results of magnetoresistivity of (GaSb)$_2$Mn and (GaSb)$_2$Cr samples in the temperature range $4-350$ K. Though in general the theory of magnetoresistance in the case of magnetic materials is very complicated and its analytical description requires solution of the Hubbard model\cite{sinjukow:prb04,arnold:prl08}, in our case we propose experimental procedure which enables us to separate the contribution of different scattering process to the overall resistivity and locate the part of it, which in our opinion is probably caused by the spin-polarized electron transport.

\section{Sample preparation and experimental details}

The samples were obtained by quenching from the synthesis temperatures in the range 700-900 K to room temperature in Toroid anvils \cite{hpr:khvostantsev2004} under the synthesis pressures in the range P=6--8 GPa. The initial mixture consists of zinc-blend GaSb and a varying content of Cr and Mn in atomic proportion 1:4   (i.e., according to chemical formula (GaSb)$_2$M. The morphology and chemical composition of the obtained samples were checked after the synthesis using a JEOL electron microscope  with the elemental microanalysis function based on the energy dispersive spectrometer. Structure determination was carried out with a Bruker diffractometer device with a Cu$K_{\alpha}^{1}$ radiation source and curved germanium monochromator. These measurements demonstarted that the samples consist practically uniformly of the single phase with point-like precipitates of surplus amounts of initial components. 

For resistivity measurements the obtained polycrystalline pellets with typical size of several millimeters were cut into samples with the thickness  of about 0.5 mm. 
Resistivity measurements were performed in the helium-free cryostat
with two-stage cryocooler (Sumitomo Cryogenics) and superconductive
magnet in the temperature range up to 300K and in the magnetic field
up to 8 Tesla. Temperature was controlled by LakeShore LS340 with
Cernox sensor with very low magnetoresistance, so that measurement
inaccuracies in the whole range of temperatures and magnetic fields
were negligible. Four-probe DC method was implemented in the
van-der-Pauw geometry with contact commutation by relays, electric
contacts to the sample were made by conductive Silver Paint. Sample
current was stabilized and measured by Keithley 2400 source meter,
voltage was measured by Keithley 2002 high precision voltmeter. 

\section{Results}

\begin{figure}
\includegraphics[width=\textwidth]{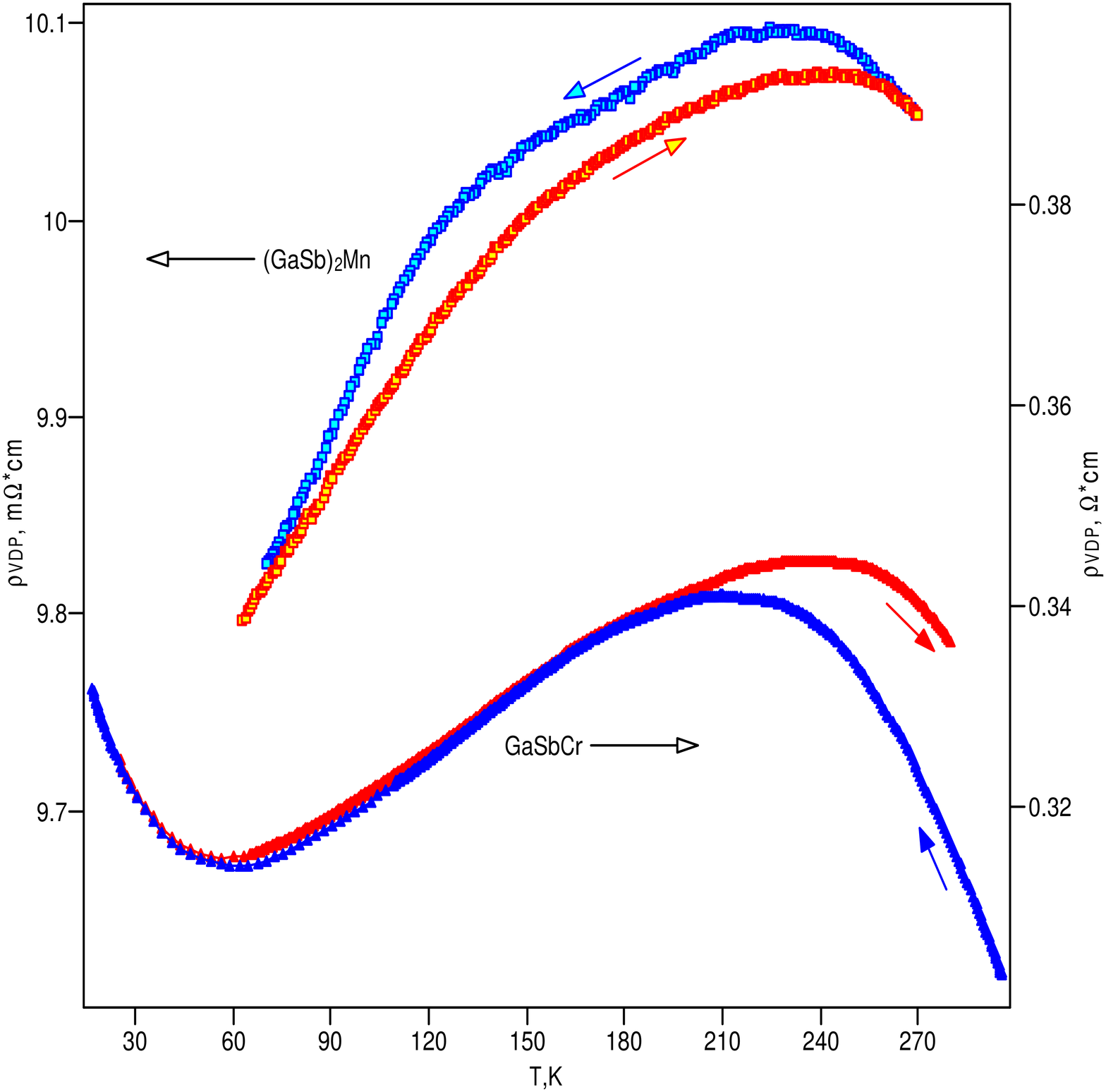}
\caption{Temperature dependence of resistivity of $(GaSb)_2Cr$ and $(GaSb)_2Mn$ samples.}
\label{fig:res}
\end{figure}

Resistivity measurements of polycrystalline samples are presented in Fig.~\ref{fig:res}. The obtained data compatible with previously published results, particularly -- the high level of resistivity $pho \approx 0.01 - 0.1 $ Ohm cm  (in GaSbCr it is almost of order of magnitude higher than in GaSbMn) and weak (almost constant) dependence of resistivity on temperature. Nonetheless at higher temperatures 250-300 K on can observe  wide maximum on temperature dependence of resistivity. 

\begin{figure}
\includegraphics[width=\textwidth]{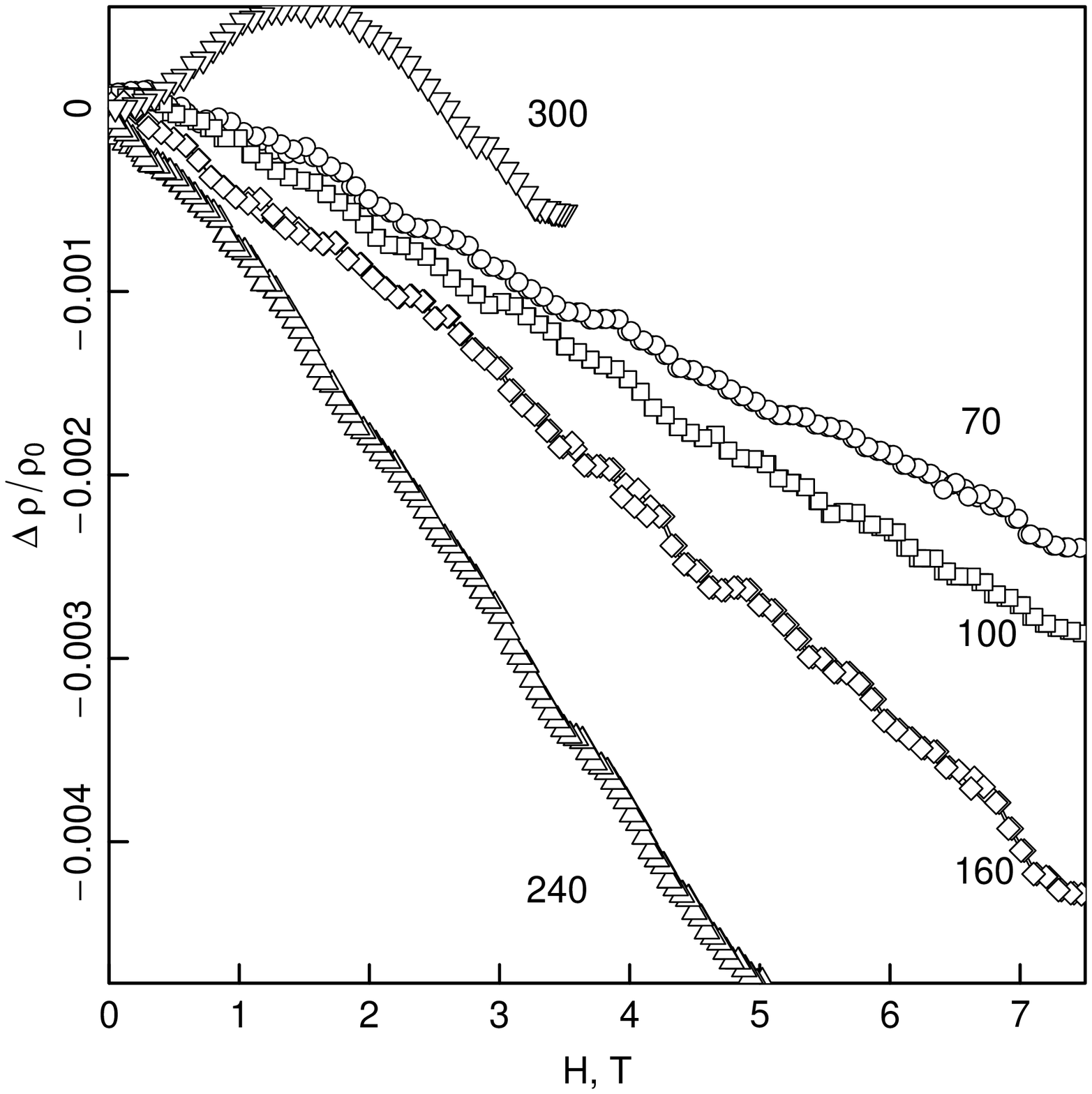}
\caption{Magnetoresistance of $(GaSb)_2Mn$ at different temperatures (temperature values are shown near respective curve). }
\label{fig:mn}
\end{figure}

\begin{figure}
\includegraphics[width=\textwidth]{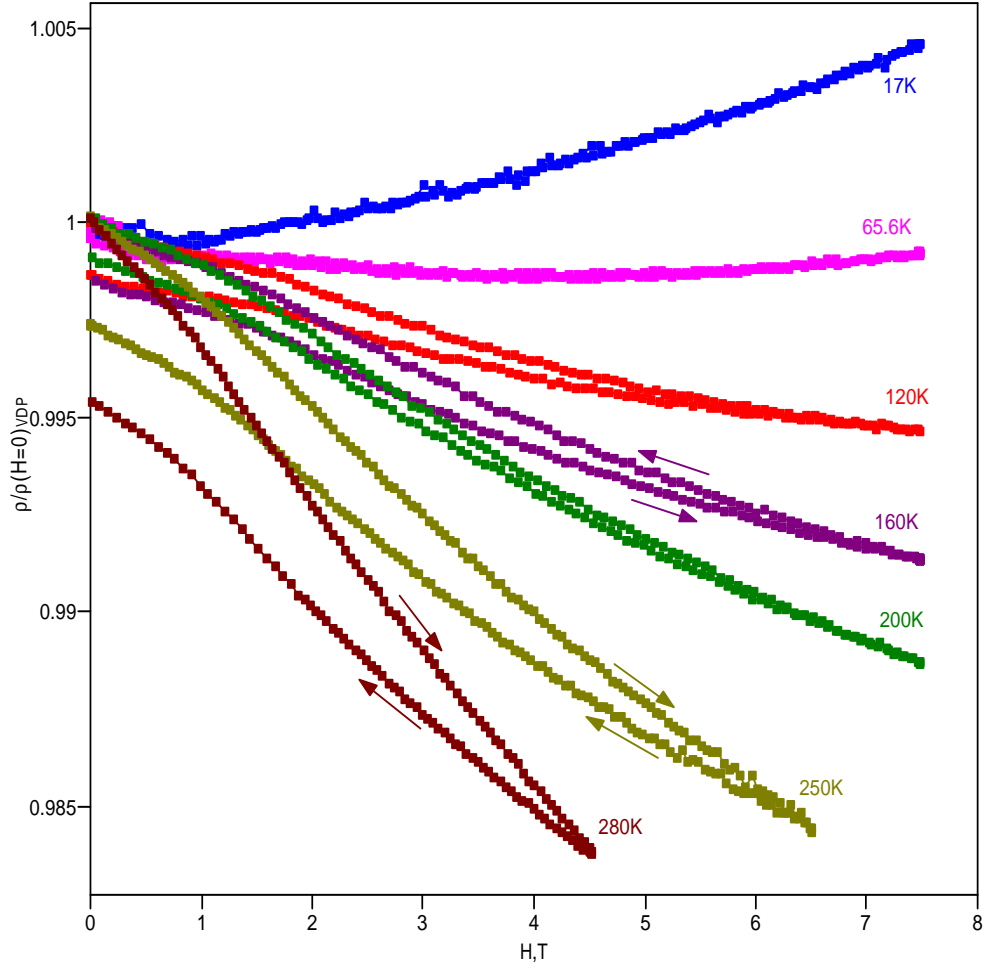}
\caption{Magnetoresistance of $(GaSb)_2Cr$ at different temperatures (temperature values are shown near respective curve) .}
\label{fig:cr}
\end{figure}

The magnetoresistance data collected at different temperatures for (GaSb)$_2$Mn and (GaSb)$_2$Cr are presented in Figs.~\ref{fig:mn} and \ref{fig:cr} respectively. Despite significant difference between these two figures, there are some common features. Firstly, in both cases in the field dependence of magnetoresistance the negative component is present and the amplitude of this component is increasing with temperature rising and amounts to 0.5-1.5 \% about room temperature. On the other hand the amplitude of the positive magnetoresistance  is decreasing with temperature rising. The positive component of magnetoresistance  is especially large in (GaSb)$_2$Cr, but it still can be discerned in the (GaSb)$_2$Mn sample too as a change in the curvature sign of experimental field dependencies in Fig.~\ref{fig:mn} collected at different temperatures (e.g. compare the curves at 60 K and 240 K). Despite the significant technical difficulties of magnetic measurements at high temperatures we collected experimental evidence that the negative component of magnetoresistance vanishes above Curie's temperatures. The curve at 300 K  in  Fig.~\ref{fig:mn} demonstrates that at critical temperature the magnetoresistance amplitude drops significantly. 

At the temperatures close to critical  the strong enough magnetic fields (for example, above 3 T in our case) can disrupt ferromagnetic ordering so different parts of the sample can be in the mixed ferromagnetic-paramagnetic state. These arguments can be applied for description of magnetoresistance hysteresis observed in  (GaSb)$_2$Cr samples (Fig.~\ref{fig:cr}). Though it was demonstarted earlier \cite{kondrin:jopcm11} that the critical temperature slightly depends on chromium content in this compounds, but one can not exclude the possibility of similar critical field dependence. However certainly the problem of observed hysteresis requires further investigation.

\section{Discussion}
It is clear that magnetoresistance in (GaSb)$_2$Mn and (GaSb)$_2$Cr consists of several components and here we will provide the procedure of their separation. The most obvious component is a positive magnetoresistance (which is relatively large in (GaSb)$_2$Cr). It is observed in wide class of materials and usually its field dependence described as a power of four of a factor between magnetic length ($l_H = \left(\frac{\hbar c}{2 e H}\right)^{1/2}$) and the length characterizing the electron transport. This last component may be a scattering length in more or less pure  metals and semiconductors \cite{en-ox*gant}, localization length in amorphous semiconductors in the hopping regime \cite{shklovskii-efros:84} or dephasing length in the heavily doped semiconductors \cite{shlimak:prb97,kondrin:jetpl08} close to the dielectric-metal transition, so-called diffusive electron transport\cite{en-ox*gant}. In all these cases it is well experimentally established that the magnetoresistance is positive and the theory produces more or less satisfactory explanation of its field and temperature dependence. Its amplitude is almost proportional to the square of magnetic field (though this dependence  can be more complicated in systems close to percolation limit in the diffusive regime\cite{shlimak:prb97,friedland:jpcm90}) with magnitude of magnetoresistance diminishing with the temperature rising. Obviously, these models can be used to describe only small part of magnetoresistance of our samples.

Another model, recently attracted considerable interest for the descriptions of magnetoresistance in the paramagnetic phase of MnSi \cite{demishev:prb12}, is the Yosida model \cite{yosida:pr57} which correlates magnetoresistance to the square of magnetization ($M$):
\begin{equation}
\frac{\Delta \rho}{\rho} \sim -a M^2
\label{yosida}
\end{equation}

Earlier this model was successfully applied to explanation of magnetotransport properties in the thin films of Fe$_3$Si \cite{vinzelberg:jap08} and InMnSb \cite{peters:prb10} and low-temperature transport properties of bulk  FeSi samples \cite{glushkov:jetp04}.  Interesting to note that all these materials are compositions of Fe-group metal and substances which in pure form have diamond-like lattices.  In all these cases the magnetoresistance is negative in accordance with the prediction of the Yosida model. 

\begin{figure}
\includegraphics[width=0.9\textwidth]{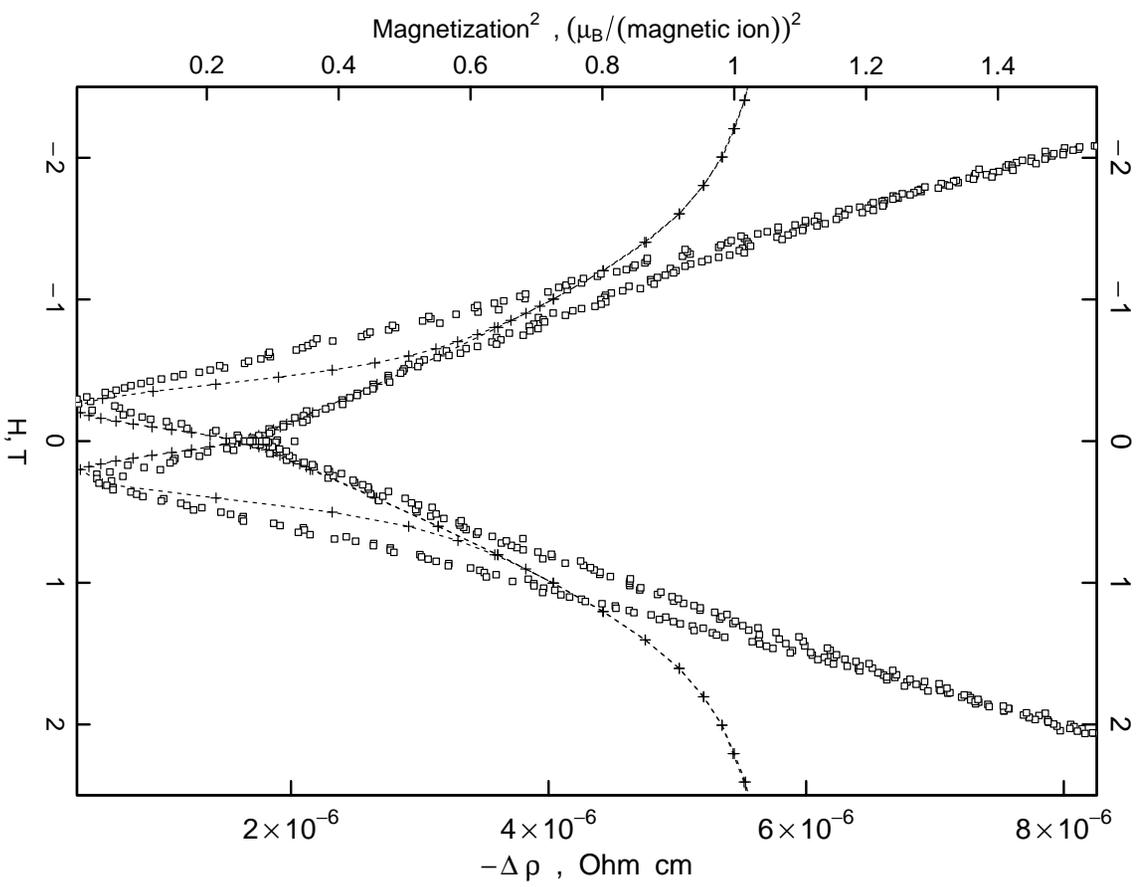}
\caption{Hysteresis of magnetisation ($+$, left axis)  and magnetoresistance ($\square$, right axis) in $(GaSb)_2Mn$ at T$=77$ K}
\label{fig:magn}
\end{figure}

\begin{figure}
\includegraphics[width=\textwidth]{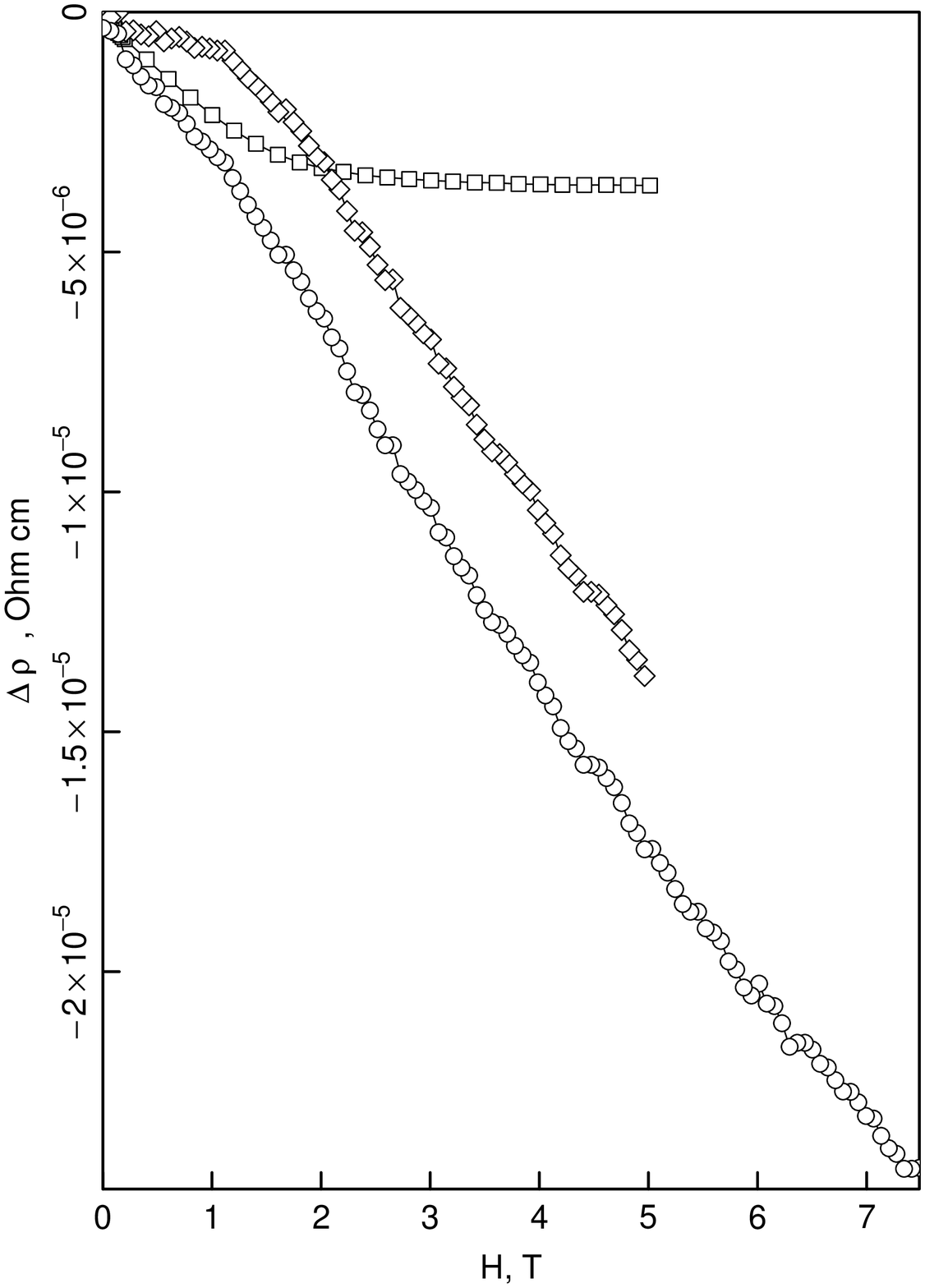}
\caption{Components of negative magnetoresistance in $(GaSb)_2Mn$ at T$=77$ K. $\circ$ -- overall magnetoresistance, $\square$ -- the Yosida component (proportional to the square of magnetization) and their difference -- the ``spin-polarized'' component ($\lozenge$).  }
\label{fig:components}
\end{figure}

However we can prove experimentally that the Yosida model is only partially applicable for description of negative magnetoresistance in (GaSb)$_2$Cr and (GaSb)$_2$Mn. For this purpose we carried out magnetoresistance measurements in sweeping magnetic field up to 2 Tesla in the way routinely used for determination of hysteresis curve in magnetic materials -- initial rise of the field (to 2 Tesla) and then sweeping it backward (to $-2$ Tesla) and then forward. Because magnetization of (GaSb)$_2$Cr and (GaSb)$_2$Mn at nitrogen temperatures characterized by considerably high coercive force ($\approx 0.2 $ Tesla) the similar hysteresis was  observed in magnetoresistance too (see Fig.~\ref{fig:magn}). The comparison with the available magnetization data provides us with the estimation of coefficient $a$ in the Yosida relation Eq.~(\ref{yosida}). In other words, this procedure amounts to  the visual match of hysteresis loops in square of magnetisation and magnetoresistance as depicted in Fig.~\ref{fig:components}. One should mark the qualitative difference in the field behavior of the magnetisation and magnetoresistance. While magnetization is saturated in the fields above 3 Tesla (so the Yosida component of magnetoresistance must saturate too), but the similar saturation of magnetoresistance wasn't observed at fields up to 8 T. So one may conclude that at the temperature T=70 K and the field 6 Tesla the Yosida model accounts only for 20\% of the magnetoresistivity (Fig.~\ref{fig:components}).

Additionally, the saturated magnetization must decrease while approaching Curie temperature and such a decreasing was actually observed in practice\cite{kondrin:jopcm11}. On the other hand, the experimentally observed magnetoresistance increases. If we took the coefficient in the Yosida model for temperature independent constant and subtracted this component from experimentally observed magnetoresistance, in the case of (GaSb)$_2$Mn we would obtain a temperature dependence of negative magnetoresistance amplitude which we consider an approximation of the component caused by the spin-polarized electron transport. In other words, the Yosida component of magnetoresistance depicted in Fig.~\ref{fig:components} at T=77 K is the evaluation from above of the contribution of this kind to the magnetoresistance at higher temperatures. So one may conclude that the relative contribution of Yosida's magnetoresistance to the total magnetoresistance of (GaSb)$_2$Mn  at 240 K drops more than twice and at 6 Tesla it should be below 5 \%. The negative magnetoresistance observed above Curie's temperature can be attributed to the Yosida mechanism due to the induced magnetisation of the samples in paramagnetic phase in a way similar to one used in the Ref.~\cite{demishev:prb12}.

Although we can not provide conclusive evidence that the physical mechanism which brings about this contribution is really caused by the spin polarization of carriers, one may mark certain similarities in the behavior of this ``spin-polarized'' component of electron transport in (GaSb)$_2$Mn and the behavior of magnetoresistance in the materials where polarization of electron transport is well established like EuO \cite{shapira:prb06}, GdN \cite{granville:prb06,leuenberger:prb05} and diluted magnetic semiconductors GaSb:Mn \cite{matsukura:02}. There were also observed negative magnetoresistance with the maximum of its amplitude in vicinity of Curie temperature.

Another possible explanation of negative magnetoresistance in (GaSb)$_2$M can be associated with weak localization effects\cite{en-ox*gant}. Influence of these sort of effects was earlier observed on the magnetoresistance of amorphous GaSb, synthesized at high pressure \cite{demishev:jetp96}. However, the preconditions for these sort of effects to be observed are significant disorder which must be present in the material and low enough temperatures and magnetic fields. In the work \cite{demishev:jetp96} negative magnetoresistance was registered only at T = 4.2 K and H $<$ 2 T.

So we can conclude that the large part of magnetoresistance of ferromagnetic phases of  (GaSb)$_2$M (M=Cr,Mn) can be attributed to spin polarized electron transport. The detected Yosida component of magnetoresistance can account for at maximum to 5 \% of magnetoresistivity (GaSb)$_2$Mn at temperatures just below Curie's temperature.

\ack 
The work has been supported by RFBR grants 13-02-00542, 13-02-01207 and 14-02-00373. The authors are grateful for N. F. Borovikov and I. P. Zibrov for X-ray and electron microscopy characterization of the samples, and thank  V. V. Brazhkin for useful discussion.

\end{document}